# Challenging Eukaryogenesis: The Story of the Eukaryotic Ancestor

Monica Vidaurri

**Abstract**
As the question of the origins of eukaryotes comes closer to an answer, the interactions between the archaeal ancestor to eukaryotes and environmental, molecular, bacterial, other influences and symbionts have become increasingly relevant. Various studies have pointed towards the Lokiarchaeota as a close archaeal relative to the eukarya. A study by Imachi et al., published in early 2020, was able to decipher the proteome and genome of the Loki and reported an entirely new model for eukaryogenesis due to its findings, thus shifting research of eukaryogenesis away from models focusing on endocytosis and the rapid acquisition of the alphaproteobacteria that would become the mitochondria. Imachi and colleagues proposed that ancestors of the Loki lived symbiotically with other prokaryotes due to its shape, and facilitated molecular transfer between the symbionts as well as the membrane manipulation necessary to allow such transfer, before acquiring a proto-mitochondria over time. This paper will lay out the history of endosymbiosis and the prevailing theory of endocytosis between archaea and alphaproteobacteria, and consolidates the knowns and unknowns up to before the publication of Imachi et al. The findings of their novel study will be discussed, and the knowns and unknowns with respect to these new findings are re-consolidated. The two models of eukaryogenesis are compared and, in moving forward with Imachi and colleagues' model, probable areas of research to further validate this approach are discussed, as well as potential usage of research tools of other fields to answer the persistent unknowns of how eukaryotes came to be.

**1. Introduction and the History of Endosymbiosis Theories**

The origin of eukaryotes remains one of the most perplexing and heavily debated topics in biology. The proverbial "missing link" in understanding how complicated eukaryota arose in an early Earth biota of relatively simple prokaryotic organisms is still largely speculative, yet wholly necessary to many biological, evolutionary, and geological questions; understanding how multicellular life arose on Earth and subsequently changed the global ecosystem and geochemical cycles even answers questions pertaining to astrobiology, or the search for life elsewhere in the universe. Namely, the study of the machine-like precision and coordination of the complex eukaryotic intracellular arrangement is largely thought to have underwent an equally complex history of evolution; one with a prokaryotic archaean ancestor. Often referred to as eukaryogenesis, the study of the evolutionary rise of eukaryotes has long been dominated by a prevailing endosymbiotic theory (Sagan 1967 and others). This theory suggests that an archaeal host cell consumed a bacterial cell – a mitochondria-like alphaproteobacteria – thus naming archaea as an ancestor to eukarya (Margulis 1996; M. W. Gray 1999; Andersson et al. 1998; Kurland 2006).

Studying the ways in which the complexity of the cellular arrangement in eukaryotes came to be also involves studying every aspect, biotic and abiotic, that influenced this event. Understanding the organisms that acted as precursors to the eukaryotes and their lineages, as well as the key environment that was needed to catalyze the makings of the eukaryote, will allow biologists to decipher exactly how eukaryogenesis happened, including how much time it took, which molecules and organisms were involved in the evolutionary transition, stages of the transition, and how the characteristics of the eukaryotic ancestor differ at each stage (Dacks et al. 2016). Widely accepted within the biology community are the events that transpired in Figure 1, which represents the general timeline of eukaryogenesis. Within this timeline lay three major transitional periods: the emergence of the Last Archaeal Common Ancestor (LACA), the First Eukaryotic Common Ancestor (FECA), and from FECA, the emergence of the Last Eukaryotic Common Ancestor (LECA) of which all eukaryotes are direct descendants.

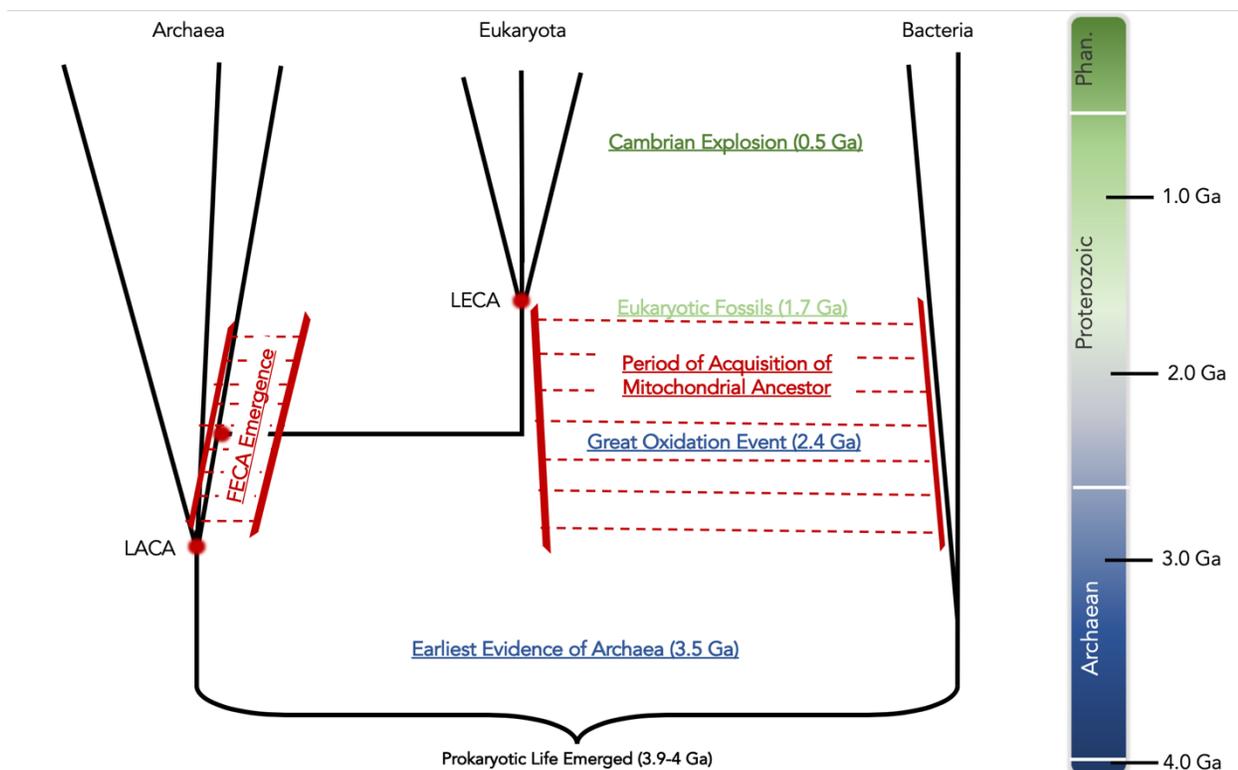

*Figure 1. General timeline of eukaryogenesis, with significant eons designating a time scale on the right with time moving from bottom to top. Ga signifies billions of years. The last universal common ancestor (LUCA) emerged roughly at the beginning of the Archean eon, with the appearance of LACA at just past 3Ga, and the appearance of LECA at roughly 1.7 or 1.8 Ga. Major milestones are also delineated, including the Great Oxidation Event and the Cambrian Explosion. The period of FECA emergence denotes the time period between LACA and LECA in which FECA acquired the alphaproteobacterial ancestor to the mitochondria: the period of acquisition of the mitochondrial ancestor, facilitated between the archaeal lineage and the bacterial lineage as the archaea obtains the bacterial cell via endocytosis.*

Numerous discrepancies of eukaryogenesis models arise from attempting to capture the monophyletic origins of eukaryotes; once a close archaeal relative to eukaryotes is identified, then the physical endocytosis of the first mitochondria can be inferred with respect to the relative's capabilities and environment. This putative ancestor does not have clear origins; as studies progress on the nature of LACA, FECA, and LECA, it becomes clear that these ancestral organisms were more complex than previously hypothesized (Margulis et al. 2006; Rochette, Brochier-Armanet, and Gouy 2014). Indeed, these three key organisms are moving targets. In the transition from LACA to FECA, these organisms had developed a membrane trafficking system, cytoskeletal proteins, and had already had the mitochondrial ancestor (Dacks et al. 2016; Gray 1999; Eugene V Koonin 2010a); though whether the mitochondria came first or last in this time frame is still widely debated (Kurland 2006).

Thus, there is a heavy concentration of study on the emergence of LECA, in which many key eukaryotic cellular functions are widely believed to have already been in place at this evolutionary stage (Dacks et al. 2016; Eugene V Koonin 2010a; Margulis et al. 2006). The existence of LECA itself is a key component in understanding the lineage of eukaryotes, and the function of actin-regulated phagocytosis as well as the phagocytic derivatives (such as motility) seen in eukaryotes are all traced back to LECA, but cannot be clearly discerned in the genetic map leading up to LECA (Koonin 2010a). Metagenomic studies of LECA give rise to two possible candidates of a eukaryotic ancestor from archaeal phyla. The first is the TACK archaeal group, which includes Korarchaeota, Thaumarchaeota, Crenarchaeota, and Eryarchaeota (Guy and Ettema 2011; Eugene V. Koonin 2010b; Koumandou et al. 2013). The second is the Asgard archaeal group, which includes Lokiarchaeota, Thorarchaeota, Odinarchaeota, and Heimdallarchaeota (Zaremba-Niedzwiedzka et al. 2017; Da Cunha et al. 2017). Both the TACK and Asgard archaeal groups, and their subsequent phyla, are now known to represent key eukaryotic signature proteins (ESPs); namely those that regulate membrane trafficking, as well as specialized proteins involved in the handling and replication of genetic material (Williams et al. 2013; E. V. Koonin and Yutin 2014; Guy and Ettema 2011). For a few years, the TACK archaea were the candidate of choice for characterizing LECA (Guy and Ettema 2011), however recent studies have pointed towards the Asgard archaea as a more viable candidate for the origins of eukaryotes (Zaremba-Niedzwiedzka et al. 2017; Fournier and Poole 2018; Spang et al. 2015; Imachi et al. 2020; Da Cunha et al. 2017; Spang et al. 2018). Studies regarding the TACK/Asgard placement in the archaeal taxonomy have formed a running interpretation of the sister relation between the two groups (Eme et al. 2017; MacLeod et al. 2019) as well as the possibility of a specific Eukaryomorpha lineage from within the archaea that facilitated eukaryogenesis and further specialized ESPs, (Fournier and Poole 2018) as shown in Figure 2.

    At this point, the author wishes to denote the central difference between FECA and LECA, and why LECA is the area of focus for many studies into eukaryogenesis given our current technological capabilities. FECA is defined as the first generation after divergence from its prokaryotic relatives, whose only living descendants are the eukaryotes, but once had prokaryotic descendants. In contrast, all eukaryotes are directly descendant from LECA as it is the youngest ancestor of extant eukaryota. Thus, FECA is inferred to be related to the last common ancestor of the Asgard, and therefore, all eukaryotes (Eme et al. 2017). It was only recently that the Asgard was named as the closest archaeal relative to eukaryotes, which makes the study of LECA more intuitive and accessible than FECA, whose characteristics are largely unknown, and no organism has been deemed a potential relative.

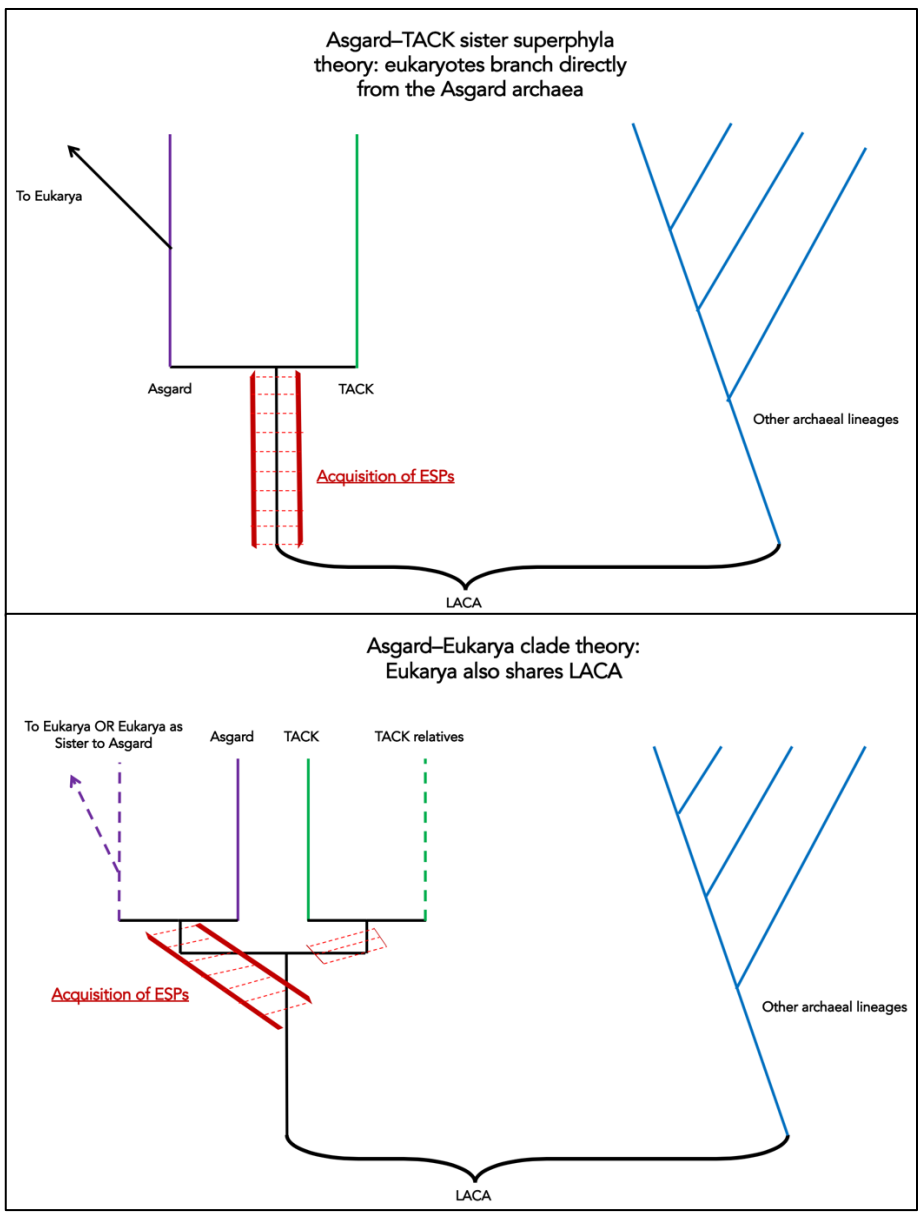

*Figure 2. A comparison of Asgard and TACK archaea as sister lineages (top) with Asgard and Eukarya as sister lineages (bottom). The top figure shows both Asgard and TACK archaeal groups as sisters, whose evolutionary appearance is marked by an earlier period of ESP acquisition. In this scenario, Eukarya splits off as a branch of the Asgard. The bottom figure is a mixed interpretation of a possible Eukaryomorpha superphylum, utilizing conclusions reached by Fournier and Poole (2018) and others. The bottom figure depicts ESPs heavily favoring the Asgard-Eukarya clade which signals the appearance of the clade itself (while TACK receives ESPs as well). Eukarya and Asgard, in this scenario, are nested as a clade of Archaea, indicating that both Eukarya and Asgard share LACA as an ancestor. In this Eukaryomorpha superphylum, the Eukarya can represent a stem itself (as denoted by Fournier and Poole (2018)), or the eukarya may have evolved from another archaeal group within the Eukaryomorpha stem that is sister to the Asgard, the likes of which is currently unknown or under speculation.*

Recently, Spang et al. (2015) named Lokiarchaeota of the Asgard archaeal group as the closest eukaryotic ancestor. Studies into both the Asgard (Villanueva et al. 2016; Zaremba-Niedzwiedzka et al. 2017) and the Loki (Embley and Williams 2015; Da Cunha et al. 2017) confirmed that this archaeal group and this phylum represents the closest known relatives of eukaryotes. This is a theory that is well-supported: of all ESPs, proteins involved in membrane modification and manipulation for endocytosis and cytoskeleton formation are crucial in the characterization of LECA because of the central feature of LECA: the acquisition and regular use of a proto-mitochondria and other molecules (Dacks and Field 2007; Bohnsack and Schleiff 2010; Yutin et al. 2009; Field and Dacks 2009). These Asgard archaea are now known to encode an abundance of ESPs, including those that are key facilitators of membrane and vesicle formation, transportation, ubiquitin and cytoskeleton formation, and others, and the Loki contain the closest known archaeal proteome to eukaryotes out of all the Asgard.

Up to this point, there have been many conclusions reached about LACA, FECA, and LECA, the characteristics at each milestone, and how and when the events of eukaryogenesis transpired. However, endosymbiosis models of eukaryogenesis present gaps in knowledge that deem the theory incomplete or insufficient. Because of these gaps in knowledge, the origins of eukaryotes and the exact timeline and key elements of eukaryogenesis remain heavily debated. Section 2 consolidates running conclusions regarding eukaryogenesis, as well as these gaps in knowledge that maintain the elusiveness of the event itself, and introduces the most novel approach to eukaryogenesis to date. Section 3 dives further into this novel study which proposes a new model of eukaryogenesis, its conclusions, and speculates whether or not the study helped resolve these gaps in knowledge mentioned in Section 2, or if it adds more uncertainty surrounding the theory of eukaryogenesis. Section 4 re-consolidates our knowledge to date, and proposes a path for further study in eukaryogenesis given the current literature, which effectively chooses between the two theories: that of endocytosis of a proto-mitochondria, and the newest proposed theory of eukaryogenesis.

**2. Prior Knowledge of Endosymbiosis and its Shortfalls**

This newest theory draws its conclusions by exploring the pure isolated genomics of the Asgard archaea, and specifically the Lokiarchaeota, which has made quite an impact in the study of eukaryogenesis. This novel study put forth by Imachi et al. (2020) introduces a new model for eukaryogenesis: entangle-engulf-endogenize (or $E^3$). Before taking into account the $E^3$ model and its findings into our knowledge of eukaryogenesis, I will consolidate the findings of generations of eukaryogenesis study up to a point prior to the publication of this groundbreaking approach.

*2.1. The "Knowns" of Eukaryogenesis*

1. <u>The Asgard archaea, and specifically the Lokiarchaeota, are the closest known prokaryotic relatives of eukaryotes.</u> Characteristics displayed across both TACK and Asgard lineages that share numerous similarities with eukarya are consolidated and displayed within Candidatus Lokiarchaeota, discovered and studied by Spang et al. (2015; 2018). The evidence for the Loki as a close eukaryotic relative is substantial. Spang et al. (2015) cites the presence of roughly 175 ESPs, which includes a ribosomal protein homologue that is heavily eukaryotic in essence. Other similarities between the Loki and eukarya are numerous and include eukaryotic cytoskeleton proteins and information processing proteins, with researchers focusing in membrane manipulation and permeation activity specific to endocytosis and molecular intracellular movement. Despite the numerous similarities, Lokiarchaeota and the Asgard group are still undoubtedly archaea, but have since been deemed to be the current front-runners in relatives of LECA itself.

2. <u>The discovery and characterization of Lokiarchaeota suggest that eukarya stemmed from archaea directly.</u> In addition to the traits described above, one of the most notable characteristics of the Loki proteome is the theorized categorization of numerous proteins associated with the BAR/IMD or I-BAR domain (Dey, Thattai, and Baum 2016) – a superfamily that, like the BAR domain, is characterized by anti-parallel dimers that allow for modification (in the process of permeation) of a cellular membrane (Stanishneva-Konovalova et al. 2016). While only strongly hypothesized, this could mean that Lokiarchaeota itself is a starting point for the evolutionary lineage of the ubiquitylation of proteins that give way to critical membrane modification pathways; within the Lokiarchaea could exist the blueprints for, or the beta version of, an endocytic pathway, and this pathway may even be external to the organism given the environment and physical characteristics of the Lokiarchaea (Dey, Thattai, and Baum 2016; Yutin et al. 2009; Baum and Baum 2014).

3. Cellular complexity is older than previously theorized. Though the Lokiarchaeota greatly increased the similarities between archaea and eukarya, these key features are seen not just in the Lokiarchaea, but many of the TACK and Asgard archaeal groups (E. V. Koonin and Yutin 2014; Yutin and Koonin 2012). This indicates a proteome across the resident phyla which encode complex cellular activities that have existed long before the endocytosis of the alphaproteobacterial endosymbiont; before LECA (Field and Dacks 2009; Yutin et al. 2009; Martijn and Ettema 2013). This provokes a further analysis of the ever-evasive FECA, a possible precursor to the Asgard archaea, which may have facilitated or even established the basis for the pathways necessary to encapsulate proteobacterial vesicles and regulate inter- and intracellular molecular transfer. Figure 3 summarizes the characteristics that mark the transition of LACA to FECA, and FECA to LECA.

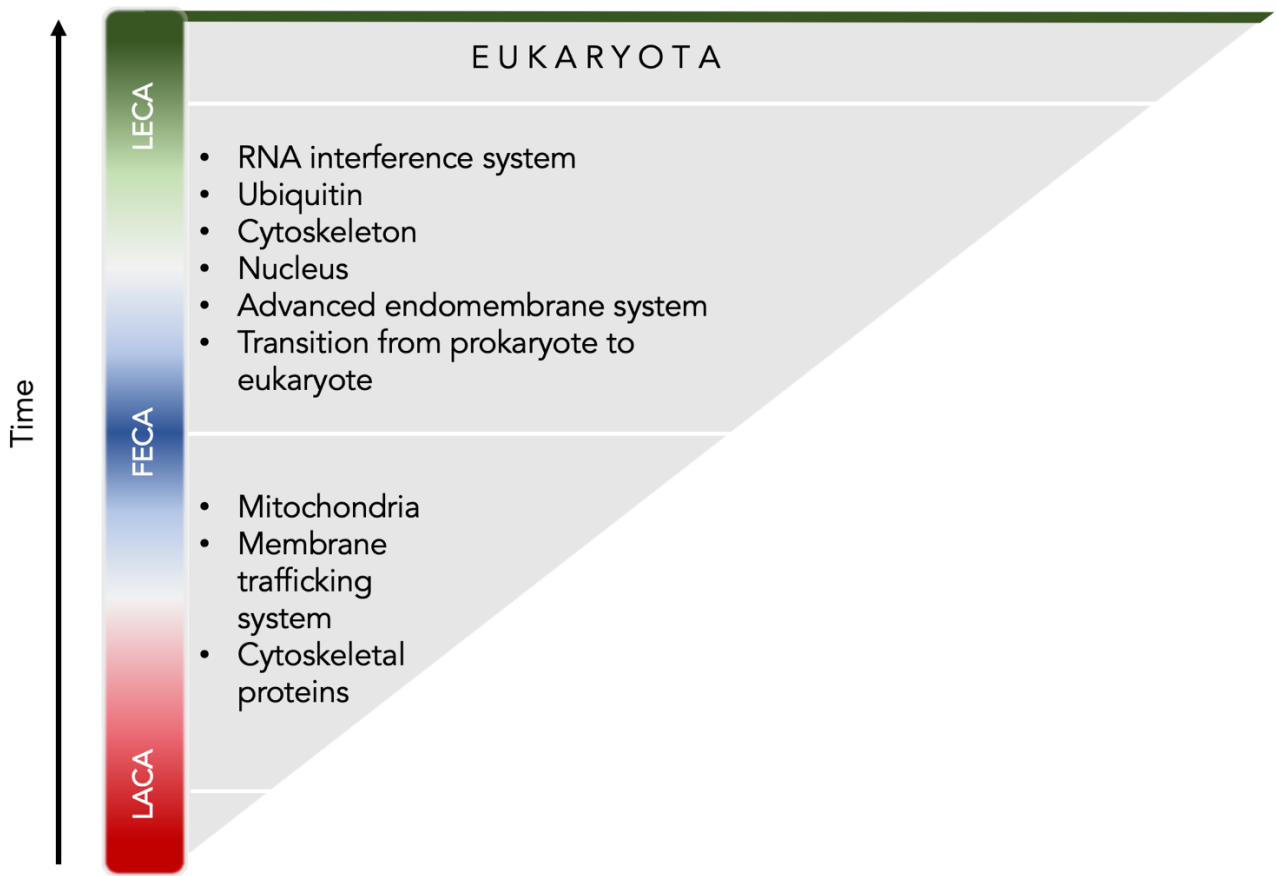

*Figure 3. Transitions between the LACA, FECA, and LECA milestones and their attributes as a function of time, which is moving bottom to top. During the transition from LACA to FECA, the LACA would have acquired a mitochondrion, membrane trafficking system, and cytoskeletal proteins, effectively signaling the evolutionary appearance of FECA. During the FECA to LECA transition, the FECA would have begun acquiring an RNA interference system, ubiquitin, a cytoskeleton, a nucleus or proto-nucleus, an advanced endomembrane system, and would have facilitated the transition from prokaryote to eukaryote, subsequently signaling the evolutionary appearance of LECA.*

4. Acquisition of alphaproteobacteria existed before LECA and after LACA (the FECA). Though it is not certain exactly when FECA emerged, studies of archaeal lineages forwards to LECA and eukaryotic lineages backwards towards LECA highlight a clear endocytic characterization of an alphaproteobacteria maximized by membrane manipulation (Poole and Gribaldo 2014). Coupled with the hypothesized complexities of LECA itself, the period of FECA emergence and evolution would likely have laid most of the groundwork for early eukaryotic cellular functions, proto-organelles and their specializations, and other key features of eukaryotes. (Gould, Garg, and Martin 2016).

However, with the discovery and findings of the Lokiarchaeota, many more questions have arisen; rather, many established questions have been repurposed.

*2.2. Questions left unanswered*

    1. The exact placement of Lokiarchaeota and Asgard. The exact period of emergence and placement of the Lokiarchaeota within the Asgard archaea is unknown, which is crucial in determining if any further studies with additional relatives is needed. In addition, further un-contaminated isolation and study of the Loki (discussed more in section 3) is necessary in discerning the role these organisms played in eukaryogenesis; they may have been directly related to the LECA organism themselves, or provide insight as to what characteristics LECA had at the time of emergence.

    2. Period of acquisition of alphaproteobacteria and its exact conditions. The question of whether the FECA endocytosis of the proto-mitochondria was the catalyst for eukaryogenesis, or the result of eukaryogenesis already taking place, is critical in studying the genomics and behavior of LECA (Knoll 2014; Poole and Gribaldo 2014; Pittis and Gabaldón 2016). Attempts to answer this question may be assisted by the existence of amitochondriate eukaryotes. Studies into the excavate *Monocercomonoides* sp. has yielded the analysis of their genomes: these amitochondriates are not only absent of mitochondria, they are absent of any mitochondria-related genes or marker proteins, yet still able to maintain its complex cellular functions without symbiosis (Karnkowska et al. 2016). Although this organism has no mitochondrion or any trace of an existence of a mitochondrion, other identifiable eukaryotic traits such as the Golgi body and genes for cytosolic pathways of metabolism were detected. Further study of amitochondriates with no trace of mitochondrial proteins may yield additional support of a late mitochondrial acquisition; it is possible that amitochondriate eukaryotes were able to split off of the FECA lineage and evolve before FECA acquired the mitochondria. Determining whether other cellular processes were present, when, and the effect it had on uptake of an early mitochondria (or vice versa) could pinpoint the time frame that LECA emerged, and the key players in the emergence.

    Another question involving acquisition of the proto-mitochondria is the act of the uptake itself. Multiple theories have been published discussing an "inside-out" method of endosymbiosis which effectively inverted the accepted event of endocytosis: these discussions propose that an ancestral archaea that is homologous to the present-day nucleus extruded membrane-bound arm-like structures that operated external to the cell and facilitated substance exchange with symbionts (Yutin et al. 2009; Baum and Baum 2014). Perhaps what is thought to be the direct endocytosis of a proto-mitochondria was instead a symbiotic relationship between an archaeon utilizing extracellular protrusions to facilitate molecular transfer with a bacterial symbiont, or multiple symbionts. This is a theory with a potentially strong lead: factors including metabolism, adaptive ability, and the changing environment in which adaptability is encouraged are known to have impacted the acquisition of the proto-mitochondria (Sojo et al. 2016), in addition to the size of the current archaeal candidate for a LECA relative. If the environment of the eukaryotic ancestor was known (beginning with further study into the hydrothermal vents where the Loki was first located), these factors may also signal the validity of these inside-out theories; low-energy environments and slow growth and molecular degradation rates will work to validate the inside-out approach.

    3. The origins of the nucleus and other organelles. Recent studies have identified evidence of archaic nuclear lamina as well as the presence of nucleoskeletal proteins in prokaryotic organisms (Devos, Gräf, and Field 2014), in addition to other prokaryotic relations to eukaryotic cellular functions and possible homologues to other eukaryotic structures, including components of aerobic metabolism and intracellular protection and maintenance (Zimorski et al. 2014; Muñoz-Gómez and Roger 2016; Leger et al. 2015). However, the origins of acquisition of other eukaryotic cellular functions and organelles is largely speculative and more elusive than the origins of the mitochondria (Gray 1989). With studies of eukaryotic ancestors emphasizing cytoskeletal formation and membrane manipulation via the early stages of a transport system regulated by membrane modification proteins, perhaps intracellular membrane modification and transport existing within the modern nucleus/endoplasmic reticulum complex act as a next step in analyzing pre-organelle functions, and the evolution of the organelles themselves. Paired with emerging studies of the chronology of endomembrane systems within a cell (namely the nuclear pore complex), this research may help solidify – or de-solidify – the running chronology of eukaryogenesis, to say nothing of the chronological placement of the mitochondria itself as the catalyst or result of eukaryogenesis. Much like the debate surrounding the pre- or post-mitochondria placement of FECA, the time period in which FECA would have existed must also have an answer to when a nucleus was acquired, as well.

    4. What characteristics did LECA have at what times, and from whom did it inherit each trait? While the latter is less intuitive given the current knowledge of eukaryogenesis, we can begin with the conclusion that LECA was much more complex and further along the evolutionary road to becoming a eukaryote than previously theorized. However, the exact level of complexity is speculation; some evidence suggests that FECA may have

already possessed the ingredients for endocytosis (Poole and Gribaldo, 2014), hinting that LECA would have been well-versed in the act. This complexity itself reaches a stalemate in the current endocytosis theory of endosymbiosis: as studies of Loki and the Asgard archaeal group emerge and are being used to derive working conclusions regarding LECA, a few issues emerge in terms of the capabilities of these archaeal candidates. These issues introduce uncertainties in the physical act of acquiring the alphaproteobacteria, and include the relatively small size of Asgard archaea compared to its suggested endosymbiont, the extent to which their low-energy allowing environment provides conditions for physical endocytosis, possible co-dependence on symbionts, and the transition from anaerobe to aerobe in its given environment (Manoharan et al. 2019). A current pre-print by Cai et al. (2018) also discusses these items as key characteristics of the Asgard.

Another question regarding complexity is the difference in membrane composition between bacteria and eukaryotes, and archaea. While bacteria and eukarya possess fatty acid chains ester-linked to a backbone comprised of glycerol 3-phosphate, archaea possess isoprene alkyl chains ether-linked to a glycerol 1-phosphate backbone. Villanueva et al. (2016) describes this as the "lipid divide", and when considering an archaeal ancestor to eukarya, requires additional study as to how this eukaryotic cellular membrane arose. This amasses uncertainty on the two-domain argument pushed by Spang et al. (2015), and raises questions regarding the early interaction between the alphaproteobacteria and the archaea; perhaps the alphaproteobacteria or another symbiont played a role in a change of membrane composition. Villanueva et al. (2016) also found that Lokiarchaeota, as well as other marine euryarchaeal groups, lack the genetic capacity to synthesize archaeal membrane lipids. Instead, these groups are able to synthesize glycerol 3-phosphate-based chimeric lipids, and may represent the stage of archaeal lipid divide within themselves. This analysis also requires further study, and the question of the lipid divide remains at large.

## 3. Novel Approaches to Eukaryogenesis

Before the $E^3$ model from Imachi et al. (2020), the Asgard archaea and its members such as Lokiarchaeota, though they were studied, had not been studied in depth, and the Loki's exact genomic characteristics were purely theorized. The team of researchers in this recent study had successfully isolated an archaeon of the Asgard archaea that is a relative of Lokiarchaeota and, due to their findings, have proposed the novel $E^3$ model of eukaryogenesis.

### *3.1. Entangle-Engulf-Endogenize*

This relative, named Candidatus *Prometheoarchaeum syntrophicum* (and denoted in their paper as MK-D1), allowed Imachi et al. to construct a ribosomal protein-based phylogenetic tree, and have found a phylogenetic sister relation between Eukarya and the Loki-relative. MK-D1 is an anaerobic coccus capable of amino acid oxidation that produces membrane vesicles and lives in a syntrophic relationship with two organisms: a hydrogen-using organism and a formate-using organism. Specific to MK-D1 symbiotes include *Halodesulfovibrio* (bacterium) and *Methanogenium* (archaea). Symbiosis with organisms is crucial to the uptake of formate and MK-D1-produced hydrogen: a high concentration of these substances was found to suppress the growth of the already ~20-day doubling cycle of MK-D1. Notably, hydrogen is a product of MK-D1 amino acid degradation, and requires other organisms to rid it of excess. 80 ESPs (also observed in the Asgard archaea) were present in this relative. Certain ESPs were observed to be highly expressed, including potential presence of proteins related to actin, ubiquitin, and ESCRT-III membrane remodeling proteins (endosomal sorting complexes required for transport, made of cytosolic protein complexes), as well as GTP-binding domain proteins. In total, MK-D1 possesses three systems that are related to cell division: actin, ESCRT, and FtsZ (filamenting temperature-sensitive mutant Z, the prokaryotic homologue of tubulin).

However, the most jarring discovery of the study of MK-D1 was its membrane-based arm-like extracellular protrusions that host its two syntropically associated organisms. Up until this study, many models of eukaryogenesis, namely the event of endosymbiosis, place hydrogen transfer between archaea and alphaproteobacteria as the mechanism that catalyzed endosymbiosis. The endosymbiosis itself is largely thought to be a physical endocytosis of a proto-mitochondria. Imachi et al. (2020) proposed that, during the Great Oxidation Event, the Lokiarchaea was instead able to use its "arms" to latch on to a suitable partner, creating a symbiotic relationship with a facultative oxygen-respiring bacterium – a theory that reflects Yutin et al. (2009) and Baum and Baum (2014) – in particularly benthic habitats of shallow oceans. This signals the "entangle" portion of the proposed eukaryogenesis model. The act of phagocytosis itself may have happened independently from endosymbiosis due to the MK-D1's low-energy output facilitated by its environment, coupled with the relatively small size of the organism. As a result, this proto-mitochondria would have facilitated the transition from syntrophy to aerobiosis via endosymbiosis; a speculation involving a biomechanical leap that researchers believe was aided by sulfate-reducing bacteria native to the environment. This signals the "engulf" portion of the model. Taking into consideration the sharing of 2-oxoacids between mitochondria and extant eukarya as well as a known behavior of Asgard archaea to obtain ATP from 2-oxoacid degradation, this engulf phase may have resulted in the host sharing a 2-oxoacid with the engulfed bacteria, to which it will have consumed the oxygen and provided the host with the amino acids, co-factors, and proteins to facilitate the transition to eukaryote. Thus, the "endogenize" portion of the model.

*3.2. Questions continued*

This study effectively narrowed down the key theories of eukaryogenesis, and some of our unknowns of pre-$E^3$ literature from Section 2.2 may re-consolidated. To begin, this study did not aim to accurately place Lokiarchaeota within the Asgard or date its emergence, and that question remains at large. This model, however, does indicate that the endosymbiosis of alphaproteobacteria may have been an ongoing process, and while the Lokiarchaea would have reaped the benefits of an oxygen-using partner during the Great Oxidation Event, the physical endocytosis and 2-oxoacid sharing between host and symbiote would have occurred later while conditions for endocytosis were working to be met, thus pointing towards late mitochondrial acquisition. This analysis also requires a closer look at the exact conditions of a related Loki ancestor in the time period of LECA. Though the exact conditions of acquisition and the question of which characteristics LECA had at which points in time remain unanswered, taking a closer look at and simulating conditions of the Great Oxidation Event can possibly signal additional characteristic development; looking at symbiotic relationships and possible entangling behavior prior to the Oxidation Event, though trivial, would answer these questions. Regardless, the results provided by Imachi and colleagues show that the $E^3$ model can be applied considerably well given the size, environment, energy output and growth, and overall capabilities of the Loki, which would have happened in a period of symbiosis before and during the Great Oxidation Event rather than immediate endocytosis.

The final component of the last unanswered question, the "lipid divide", was briefly mentioned by Imachi and colleagues. Researchers speculate that the archaeal host, like MK-D1, possessed ether-linked lipids typically characterized by the Asgard group and archaea as a domain. While the divide itself is still unanswered in this study but remains a topic of consideration, researchers note that there have been multiple suggestions put forth in recent literature regarding the evolutionary considerations of the lipid divide, and which challenge how stark of a "divide" it really is. Ester and ether lipid links may have been able to "mix" with one another without sacrificing integrity of the host membrane itself during eukaryogenesis (Caforio et al. 2018). The transition of ether to ester has also been the topic of other studies, which suggest that the transition came in parts, but point to a "hybrid" lipid biosynthesis transitional period, or rather, a "divide" that is not as stark as previously thought, and one helped along by the shared polar head groups across all kingdoms (Koga and Morii 2007).

**4. Discussion and Further Research Prospects and Partnerships**

The introduction of the $E^3$ model signaled an exciting directional shift in the study of eukaryogenesis, and one that leaves plenty of compelling new avenues for research. However, this approach is still a theory, similar to the popular endocytosis theory that predated the predictions and publication of the $E^3$ model. Given the questions answered by the $E^3$ model and questions left unanswered by it, which of these models is more "correct", or worth pursuing? Convoluted as that question may be, the answer revolves around the Lokiarchaeota and the Asgard archaeal group. Both theories point towards the Asgard archaea, and specifically the Lokiarchaeota, as a dependable starting point of eukaryogenesis study and thus a potential relative to LECA. We now know the nature of the Lokiarchaeota: small, slow in growth, anaerobic and with membrane vesicles, living in extreme environments, sporting extracellular protrusions, and dependent on symbionts that regularly interact with these protrusions. Given this information, it is unlikely that endocytosis occurred in a rapid time frame, but rather, the organism would have needed time to formulate a response to the Great Oxidation Event by choosing an endosymbiont in its available environment to help it adjust to the ordeal, existing with this symbiont for a time, and then beginning to engage in endocytosis. Imachi and colleagues also suggest that many archaeal lineages share this ability to degrade amino acids syntropically; even if the Loki turns out to be insufficient in characterizing the LECA, the suggestion that the Asgard ancestor – and archaea as a whole – share this characteristic pushes the $E^3$ theory further along to validity.

Additional study is, of course, needed to corroborate these amassing suggestions. On the basis of an $E^3$ approach to eukaryogenesis, what happens next? Now that isolation of a pure Loki-related genome is possible, perhaps all the members of the Asgard archaea can be observed for similar characteristics, or characteristics that point to another organism residing in this clade is even closer to a eukaryotic relative. Contrarily, the existence of a third party between archaea and eukarya, such as Eukaryomorpha or proto-eukarya, though not blatantly suggested in this new study, also cannot be deemed insufficient or unlikely; MK-D1 is not necessarily an "intermediate" between archaea and eukaryote, but fully archaeal. However, the intricacies of discovering such an intermediate is not immediately perspicuous: by investigating the newest clues that have become particularly notable about the Asgard and Loki and their life of symbiosis, studies of the existence of the intermediate may be either cast aside or become of higher importance. The only way we will know is to begin.

The Loki, aptly named after the mythological trickster god himself, may be an elusive character, but these characteristics themselves may be guide posts for various approaches to further study. Luckily for biologists, as our understanding of the Loki and the $E^3$ method becomes increasingly enhanced, so too does the technology required to study the organisms and the mechanisms involved. Overall, molecular dating, mass spectrometry (MS), nuclear magnetic resonance spectroscopy (NMRS), and various forms of modeling will likely represent the major approaches moving forward with validating an $E^3$ model of eukaryogenesis.

To address Loki's emergence within the Asgard and other emergences of importance, efforts largely including biogeochemists and paleontologists are being made to improve molecular dating. One such recent improvement comes in the form of Bayesian relaxed-molecular clocks using fossil data, which can pinpoint the dates of divergence and intervals at which they occur (Eme et al. 2014). Because this technique is fossil-calibrated and reliant on input of an organism's proteins and general taxa, as more fossil data comes forward from the Proterozoic eon and as more information about the genomics of the Asgard archaea become clear, the ages and divergence points of each phylum can be estimated with greater precision.

Another area of technological improvement running parallel to breakthroughs in eukaryogenesis is MS. Various developments in MS have allowed researchers to use high-mass resolution systems in order to focus in on the topic of archaeal molecular usage from complex environmental samples. Imachi et al. (2020) used gas chromatography MS and ion MS to study lipids and amino acid utilization, respectively. Strides have been made using MS in lipidomics, which studies lipid functions of various organisms both within the organism and in relation to its biome; lipidomics has direct implications on membrane manipulation and evolution, which remains a critical yet obscure component in eukaryogenesis. As a result, new studies in archaea-symbiont and archaea-environment interactions have been made emphasizing lipid membrane manipulation, utilizing various forms of MS to analyze such interactions can generate archaeal lipid structural libraries, where various interactions with differing environments, symbionts, macromolecules, and other organisms can be compared and contrasted (Law and Zhang 2019). This allows direct studies on the lipidome of the Loki and other Asgard with respect to various influences, and can be carried out using current technology.

In addition to gas chromatography MS, ion MS, and various other flavors of MS, macromolecular interactions between ancestral archaea and their symbionts, including protein-protein and amino acid interactions, can be studied using NMRS: a method used widely by biochemists and biogeochemists that can yield atomic resolution of such interactions and molecular structures themselves (Burz et al. 2018; Karl et al. 2011). Karl et al. (2011) also highlights the various uses of NMRS for applications including real-time enzymatic reactions, proteins under intracellular environmental conditions, study of nucleic acids, and others. Utilizing NMRS with MK-D1 under lab-created Proterozoic and Archean environments may elicit the proper responses taken by the eukaryotic ancestor itself on the macromolecular level during the transition to eukarya: perhaps a simulated Great Oxidation Event involving Loki within a simulated Archean environment could uncover methodology used by Loki to survive the event specific to its nucleic acids, proteome, and other eukaryogenesis-critical molecular components.

Constructing lab-created environments for pre-, during, and post-Great Oxidation Event and constructing models of these events can be used in conjunction with molecular dating, MS, and NMRS data, and can work to elucidate characteristics of ancestral eukarya at each stage and possible environmental compensation techniques. Planetary scientists, for example, already work with such models and simulations which work to imitate conditions of specific time periods of the Earth's history.

**5. Conclusion**

Delving deeper into the mechanisms of eukaryogenesis is becoming increasingly synonymous with studying the Lokiarchaeota in greater depth. In addition, the arrival of the $E^3$ model of eukaryogenesis leaves a novel approach, and one that is quite distinct from the previous theory of endocytosis that is mitochondria-centered. This difference in approach is worth pursuing given the physical characteristics of the Loki and its environment, and is not short of exciting avenues of research. With the emergence of new studies that continue to verify the proximity of relatedness between Lokiarchaeota and eukarya, studies into the environmental conditions and subsequent actions taken by Loki will provide a test of validity of the $E^3$ model of eukaryogenesis, which opens the door to collaboration with biogeochemists, paleontologists, and even astrobiologists. Each of these fields represents a technological and theoretical readiness to begin to uncover the mysteries of eukaryogenesis as it exists through the lens of the $E^3$ model. The story of eukaryogenesis is one that is rich in history and complexity of theory, much like the phenomena itself. I argue that, with new avenues into collaborations between biologists and others, the story of eukaryogenesis as told by archaeal and bacterial symbiosis is just starting to be written.